\documentclass[global]{svjour}
\usepackage{natbib}
\usepackage{graphics}
\usepackage{latexsym}
\usepackage{amsmath}
\usepackage{amssymb}
\usepackage{mathrsfs}

\newcommand{\EEE}{\boldsymbol E}
\newcommand{\DDD}{\boldsymbol D}

\journalname{Optical and Quantum Electronics}

\begin{document}

\title{Slow-light enhanced optical detection in liquid-infiltrated photonic crystals\thanks{Paper accepted for the "Special Issue OWTNM 2007" edited by A. Lavrinenko and P.~J. Roberts.}}

\author{M.~E.~V.~Pedersen \and L.~S.~Rish{\o}j \and H.~Steffensen \and S.~Xiao \and N.~A.~Mortensen\thanks{email: nam@mic.dtu.dk}}

\institute{MIC -- Department of Micro and Nanotechnology,
Nano$\bullet$DTU, Technical University of Denmark, {\O}rsteds
Plads, DTU-building 345 east, DK-2800 Kongens Lyngby, Denmark. }
\date{Received: \today / Revised version: date}
\maketitle
\begin{abstract}
Slow-light enhanced optical detection in liquid-infiltrated
photonic crystals is theoretically studied. Using a
scattering-matrix approach and the Wigner--Smith delay time
concept, we show that optical absorbance benefits both from
slow-light phenomena as well as a high filling factor of the
energy residing in the liquid. Utilizing strongly dispersive
photonic crystal structures, we numerically demonstrate how
liquid-infiltrated photonic crystals facilitate enhanced
light-matter interactions, by potentially up to an order of
magnitude. The proposed concept provides strong opportunities for
improving existing miniaturized absorbance cells for optical
detection in lab-on-a-chip systems.
\end{abstract}

\section{Introduction}
Optical techniques are finding widespread use in chemical and
bio-chemical analysis, such as absorbance, fluorescence,
Raman-scattering, and surface-plasmon-resonance measurements, see
\cite{Mogensen:2004,Hess2002,Kneipp:2002,Homola:1999}. In
particular, the Beer--Lambert absorbance measurement has become
one of the classical quantitative work\-horses in analytical
chemistry, \cite{Skoog:1997}. The increasing emphasis on
miniaturization of chemical analysis systems during the past
decade, \cite{Janasek:2006}, has naturally stimulated a
considerable effort in integrating microfluidics,
\cite{Squires:05,Whitesides:2006}, and optics in lab-on-a-chip
microsystems, \cite{Verpoorte:2003,Mortensen:2007b}, and more
recently this has to a large extent powered the emerging field of
optofluidics, see \cite{Psaltis:2006,Monat:2007a}. However,
lab-on-a-chip implementations of the above mentioned optical
techniques call for new approaches going beyond a simple
miniaturization of existing optical setups. The need for a new
perspective is perhaps best appreciated by emphasizing the optical
drawback of miniaturization. Reducing the optical path length will
eventually also decrease the sensitivity significantly. In other
words, the sensitivity fundamentally scales with volume hosting
the interaction between light and the solution of analytes.

\begin{figure}[t!]
\resizebox{\textwidth}{!}{\includegraphics{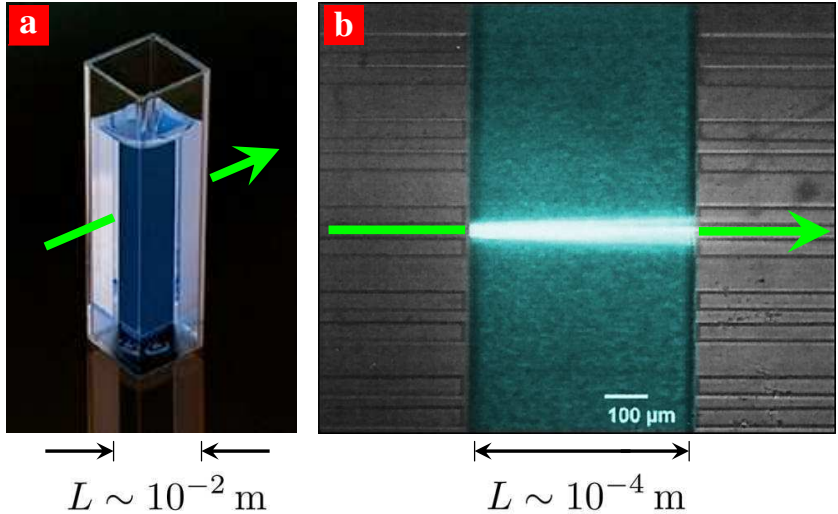}}
\caption{Panel (a) illustrates a typical macroscopic cuvette while
panel (b) shows a microscope image (top-view) of an equivalent
lab-on-a-chip implementation of a microfluidic channel (vertical
direction) integrated with planar optical waveguides (horizontal
direction). Courtesy of K.~B. Mogensen and J.~P. Kutter (MIC --
Department of Micro and Nanotechnology, Technical University of
Denmark, www.mic.dtu.dk).} \label{fig1}
\end{figure}

Panel (b) in Fig.~\ref{fig1} illustrates a typical lab-on-a-chip
implementation of an absorbance cell. The optical path length $L$
is often reduced by several orders of magnitude compared to
typical macroscopic counterparts, such as the cuvette shown in
panel (a). A typical size reduction by two orders of magnitude
will penalize the optical sensitivity in an inversely proportional
manner, \cite{Mogensen:2003}.

In the above context, photonic crystals (PhC) are potential
candidates for the desired new approach on the light-matter
interactions as recently reviewed by \cite{Mortensen:2007b}. The
concept of PhCs refers to a class of artificial electromagnetic
structures offering highly engineered dispersion properties, as
first suggested by \cite{Yablonovitch:1987,John:1987}. PhCs are
porous structures, but with the voids arranged in a highly regular
way. Thus, for a wavelength $\lambda$ comparable to the
periodicity $\Lambda$ of the PhC, the electromagnetic radiation
interacts strongly with the matter. PhCs have over the years been
recognized as strongly dispersive environments supporting a number
of phenomena including photonic band gaps and slow-light
propagation, see \cite{Joannopoulos:1995,Sakoda:2005}. The
key-point in the present context is that liquid-infiltrated PhCs
will overall inherit the unusual dispersion properties from their
non-infiltrated counterparts, thus also changing the way light
interacts with bio-molecules dissolved in the liquid,
\cite{Mortensen:2007b}. The effects can be quite pronounced
compared to light-matter interactions in a spatially homogeneous
liquid environment and in this paper we follow this route to
compensate the reduced optical path.

In particular, we consider slow-light enhanced optical detection
in liquid-infiltrated photonic crystals, thus potentially
compensating for the cost of miniaturization and reduction in
optical path length.

The enhancement factor $\gamma\equiv \alpha/\alpha_l$ for an
absorbing liquid (with absorption coefficient $\alpha_l$) can be
expressed as
\begin{equation}
\label{eq:gamma} \gamma =f\times\frac{c/n_l}{v_g},
\end{equation}
as recently derived by \cite{Mortensen:2007a}. Here, $0<f<1$ is a
dimensionless number quantifying the relative optical overlap with
the liquid. The fraction on the right-hand side expresses the
ratio of the group velocity $c/n_l$ in the bare liquid to the
group velocity $v_g$ in liquid-infiltrated photonic crystal, thus
clearly illustrating the enhancement by slow-light propagation
($v_g\ll c$). In the following, we first derive the above
expression before we with numerical simulations illustrate the
slow-light enhancement in liquid-infiltrated PhCs with three
different periodic structures.

\section{Theory}

\subsection{Electromagnetic wave equation}
The electromagnetic problem that we consider is governed by the
electromagnetic wave equation for the electrical field,
\begin{equation}\label{eq:wave}
\nabla\times\nabla\times \big|\EEE\big>=\varepsilon
\frac{\omega^2}{c^2}\big|\EEE\big>,
\end{equation}
where for the absorbing liquid the dielectric function
$\varepsilon$ has a small imaginary part, $\omega$ is the angular
frequency, and $c$ is the velocity of light in vacuum. In a recent
paper, \cite{Mortensen:2007a} derived Eq.~(\ref{eq:gamma})
rigorously with the aid of standard first-order electromagnetic
perturbation theory
\begin{equation}\label{eq:perturbation}
\delta\omega=-\frac{\omega}{2}
\frac{\big<\EEE\big|\delta\varepsilon\big|\EEE
\big>}{\big<\EEE\big|\varepsilon\big|\EEE\big>}
\end{equation}
and more details may be found in our recent work, see
\cite{Mortensen:2007b}. In Eq.~(\ref{eq:perturbation}),
$\delta\omega$ is the first-order frequency shift introduced by a
perturbation $\delta\varepsilon$ of the dielectric function
$\varepsilon$ (assumed frequency independent) with the electric
field being the unperturbed one. In the present context the
perturbation is associated with absorption in the liquid-part of
space thus giving rise to an imaginary frequency shift. In the
following we emphasize this by explicitly writing the shift as
$i\delta\omega$ with $\delta\omega$ being real. For the temporal
dependence of the intensity we have $I(t)=I_0 \big|e^{-i\omega
t}\big|^2=e^{-2\delta\omega t }$ so that we may define an
absorption rate given by $\Gamma=2\delta\omega$.

\subsection{Scattering-matrix approach}

Here, we will emphasize the light-matter interaction time and
offer an alternative derivation based on scattering matrices
combined with the concept of the Wigner--Smith delay time.
Following \cite{Beenakker:2001} we start from the scattering
matrix
\begin{equation}
S(\omega)=\begin{pmatrix}S_{11}(\omega)&
S_{12}(\omega)\\S_{21}(\omega)&S_{22}(\omega)\end{pmatrix}=\begin{pmatrix}r(\omega)&
t'(\omega)\\t(\omega)&r'(\omega)\end{pmatrix}
\end{equation}
where $S_{ij}$ are the S-parameters connecting incident and
out-going plane-wave (or Bloch) states to the left and the right
of the scattering region. Next, let $S_0(\omega)$ denote the
scattering matrix in the absence of absorption so that for a weak
absorption we have a small imaginary shift $i\delta\omega$ in
frequency and thus $S(\omega)\simeq S_0(\omega+i\delta\omega)$.
Taylor expanding the right-hand side we get
\begin{equation}\label{eq:SS0}
S(\omega) \simeq S_0(\omega)+i\delta\omega \frac{\partial
S_0(\omega)}{\partial\omega} = S_0(\omega)\left[1 -
\frac{\Gamma}{2} Q_0(\omega)\right].
\end{equation}
In the second equality $\Gamma=2\delta\omega$ is the absorption
rate and we have furthermore used the unitarity of the scattering
matrix, $S_0^\dagger S_0=1$, to introduce the Hermitian
Wigner--Smith delay time matrix
\begin{equation}
Q_0(\omega)=-iS_0^\dagger(\omega)\frac{\partial
S_0(\omega)}{\partial\omega}.
\end{equation}
Now, we would like to calculate the transmission probability $T$
through a single eigenmode in which case $S$ is a two-by-two
matrix. Assuming time-reversal symmetry we have
$T=\big|S_{12}\big|^2=\big|S_{21}\big|^2$ which we have the
freedom to rewrite as
\begin{equation}
T(\omega)=\frac{1}{2}{\rm Tr}
\left\{S(\omega)S^\dagger(\omega)\right\}+{\cal O}(1-T).
\end{equation}
For near-resonance transmission, i.e. for a low reflection, we get
\begin{equation}\label{eq:T1}
T(\omega)\simeq\frac{1}{2}{\rm Tr} \left[1 - \Gamma
Q_0(\omega)\right]= \frac{1}{2} \sum_{n=1}^{2}\left[1 - \Gamma
t_n(\omega)\right]=1 - \Gamma\tau
\end{equation}
where we have used the unitarity of $S_0$ together with the cyclic
invariance of the trace. In the second equality we have introduced
the eigenvalues $t_1=t_2=\tau$ of the Wigner--Smith delay time
matrix where $\tau$ is often referred to as the dwell time. We may
use perturbation theory, Eq.~(\ref{eq:perturbation}), to calculate
the absorption rate which gives
\begin{equation}
\Gamma=f\times \Gamma_l,\quad f\equiv
\frac{\big<\EEE\big|\DDD\big>_l}{\big<\EEE \big|\DDD\big>}
\end{equation}
where we have introduced the displacement field
$\big|D\big>=\varepsilon \big|E\big>$ and $\Gamma_l$ is the
absorption rate of the liquid itself. For the filling factor we
have $0<f<1$ and the integral in the numerator is restricted to
the region containing the absorbing fluid while the integral in
the denominator is spatially unrestricted. The presence of the
filling factor $f$ is quite intuitive since only the fraction $f$
of the light residing in the fluid can be subject to absorption.
Note that compared to the definition normally used by the photonic
crystal community, \cite{Joannopoulos:1995}, our definition
quantifies the relative overlap with the liquid-part of space
rather than the dielectric part given by $1-f$. Finally, we
rewrite Eq.~(\ref{eq:T1}) by re-exponentiating, i.e.
$(1-x)\longrightarrow \exp(-x)$,  so that
\begin{equation}
T=\exp(-\Gamma\tau)\equiv \exp(-\gamma \times \Gamma_l \tau_l)
\end{equation}
corresponding to the expected exponential damping of the resonance
due to absorption. In the second equality we have introduced
$\gamma\equiv\alpha/\alpha_l=(\Gamma\tau)/(\Gamma_l\tau_l)$ and
using that $\tau=L/v_g$ for the periodic medium and
$\tau_l=L/(c/n_l)$ in the bare liquid we finally arrive at
Eq.~(\ref{eq:gamma}).

The expression for $\gamma$ clearly demonstrates how optical
absorbance benefits from slow-light phenomena. The effective
enhancement of the absorbance may thus naturally be interpreted as
an effective enhancement of the light-matter interaction time.
These conclusions may be extended to also non-periodic systems,
including enhanced absorbance in disordered systems as well as
intra-cavity absorbance configurations, see
\cite{Mortensen:2007b}.

\begin{figure}[t!]
\resizebox{0.8\textwidth}{!}{\includegraphics{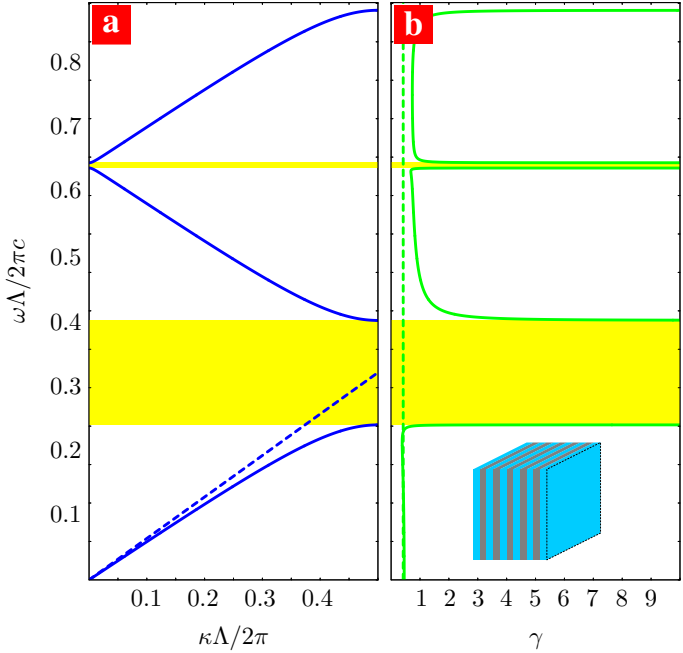}}
\caption{(a) Photonic band structure for normal incidence of
either TE or TM polarized light on a Bragg stack of period
$\Lambda=a_l+a_2$ with $n_l=1.33$, $n_2= 3$, $a_l=0.6863 \Lambda$,
and $a_2 = 0.3137 \Lambda$. Photonic band gaps are indicated by
yellow shading and the dashed line indicates the long-wavelength
asymptotic limit, Eq.~(\ref{eq:Bragg1}). (b) Corresponding
enhancement factor which peaks and exceeds unity close to the
photonic band-gap edges. The dashed line indicates the
long-wavelength asymptotic limit, Eq.~(\ref{eq:Bragg2}).}
\label{fig2}
\end{figure}

In general, fluids/liquids have a smaller refractive index than
the typical solid materials comprising the photonic crystal
itself. Thus, increasing the filling factor (and thereby the
light-liquid interaction) implies the confinement or guidance of
light in the low index regions, which is typically not the case
explored in more ordinary applications of PhCs for e.g. optical
communication purposes. However, a high light-liquid overlap can
be achieved in some structures, e.g., slot waveguides, hollow
anti-resonant reflective optical waveguides, as well as
photonic-crystal structures. For liquid-infiltrated photonic
crystals and photonic crystal waveguides, it is possible to
achieve $v_g\ll c$ and at the same time have a filling factor of
the order unity, $f\sim 1$, whereby significant enhancement
factors become feasible.

\begin{figure}[t!]
\begin{center}
\resizebox{0.8\textwidth}{!}{\includegraphics{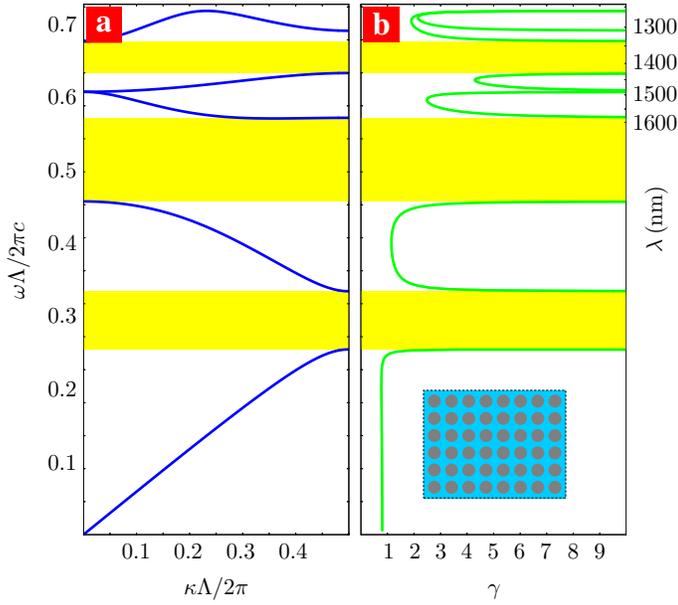}}
\end{center}
\caption{(a) Photonic band structure for propagation of TM
polarized light along the $\Gamma$X direction in a square lattice
of period $\Lambda$ with dielectric rods of diameter
$d=0.2906\Lambda$, $n=3.24$ and $n_l=1.33$. Photonic band gaps are
indicated by yellow shading. (b) Corresponding enhancement factor
which exceeds unity for the flat bands in general and the fourth
band in particular. The right $y$-axis shows the results in terms
of the free-space wavelength when results are scaled to a
structure with $\Lambda=1000\,{\rm nm}$. } \label{fig3}
\end{figure}

\section{Numerical examples}

Having derived a formal expression for the enhancement factor
$\gamma$, we next apply a full-vectorial plane wave method of
\cite{Johnson:2001} to calculate the eigenmodes of
Eq.~(\ref{eq:wave}) in the non-absorbing limit and from these
solutions we evaluate the enhancement factor,
Eq.~(\ref{eq:gamma}), for a number of geometries.

\subsection{Bragg stack}
Let us first illustrate the slow-light enhancement for the
simplest possible structure which is a Bragg stack with normal
incidence of electromagnetic radiation, as shown in the inset of
panel (b) of Fig.~2. The structure is composed by the low-index
material layers of width $a_l=0.6863 \Lambda$ being a liquid with
refractive index $n_l=1.33$ and the high-index layers have a width
$a_2 = 0.3137 \Lambda$ and a refractive index $n_2= 3$. Panel (a)
of Fig.~2 shows the corresponding band structures, where photonic
band gaps are indicated by yellow shading and the dashed line
indicates the long-wavelength asymptotic limit to a
close-to-linear dispersion
\begin{equation}\label{eq:Bragg1}
\omega(\kappa)\simeq \frac{c\kappa\Lambda}{a_ln_l+a_2n_2}.
\end{equation}
When approaching the band-gap edges, the dispersion flattens. Note
that the first band is a dielectric-like one, see e.g.
~\cite{Mortensen:2007b,Joannopoulos:1995}, meaning that most of
the energy localizes in the high dielectric region. For the first
band it is well-known that the flat dispersion originates from a
spatial localization of the field onto the high-index layers and
thus $f\ll 1$ near the band edges where the inverse group velocity
diverges. However, in spite of the localization, the enhancement
factor may still exceed unity as shown in panel (b) where the
dashed line indicates the long-wavelength asymptotic limit with
\begin{equation}\label{eq:Bragg2}
f\simeq \frac{
\big<1\big|\epsilon\big|1\big>_l}{\big<1\big|\epsilon\big|1\big>}=\frac{a_ln_l^2}{a_l
n_l^2+a_2 n_2^2}.
\end{equation}
For the second band, the enhancement factor still exceeds unity,
which is attributed both to slow-light, as well as to the large
filling factor. The filling factor is about $20\%$ at the band
edge for the second band. In order to further benefit from the
slow-light enhanced light-matter interaction, we obviously have to
pursue optofluidic structures supporting both low group velocity
and at the same time large filling factors.

\begin{figure}[t!]
\begin{center}
\resizebox{0.8\textwidth}{!}{\includegraphics{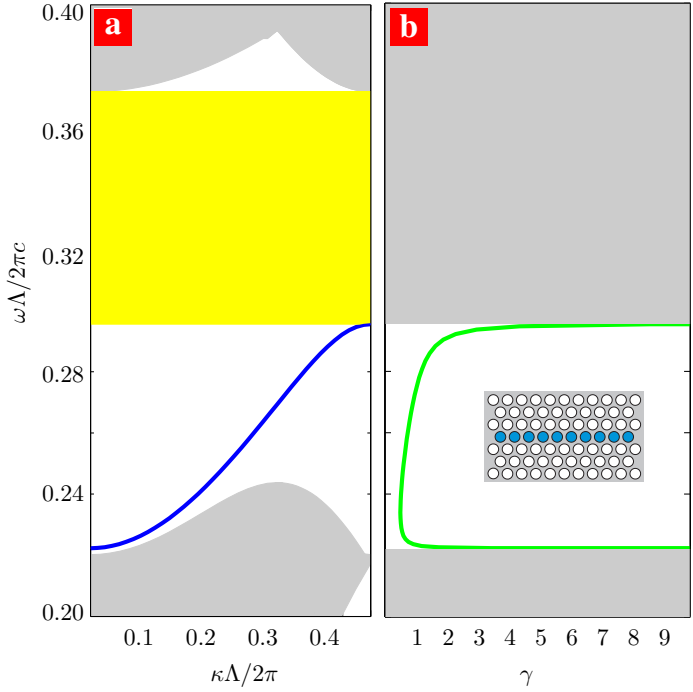}}
\end{center}
\caption{(a) Photonic band structure for propagation of TE
polarized light along the $\rm \Gamma$X direction in a PhC
waveguide in a triangular lattice of period $\Lambda$ with holes
of radius $r=0.37\Lambda$ and $n=3.24$. The PhC waveguide is
formed by enlarging the radius ($r_1=0.50\Lambda$) of the air
holes in a single row, which are infiltrated by a liquid with
$n_l=1.33$. The grey shading indicates the regions of finite
density of states in the surrounding PhC while the yellow shading
shows the waveguide mode-gap region. (b) Corresponding enhancement
factor $\gamma$. The inset shows the structure of the photonic
crystal waveguide.} \label{fig4}
\end{figure}

\subsection{Pillar-type photonic crystal}

Figure~3 shows the case for a 2D square photonic crystal with
high-index dielectric rods in liquid. Compared to the Bragg stack
in Fig.~\ref{fig2}, some of the modes in this structure have both
a low group velocity and at the same time a reasonable value of
the filling factor $f$. In particular, the fourth band in panel
(a) is quite flat. The corresponding enhancement factor $\gamma$
exceeds 4 for a bandwidth of order 100~nm for a pitch around
$\Lambda\sim 1000\,{\rm nm}$, as indicated on the right $y$-axis
in panel (b).

\subsection{Void-type photonic crystal waveguide}

Recently, it was shown by \cite{Gersen:2005} how photonic crystal
waveguides can be used to slow down the propagation velocity of
light. Here, we illustrate how this may be utilized for slow-light
enhancement in a PhC waveguide structure. We consider the PhC
waveguide shown in the inset in panel (b) of Fig.~\ref{fig4},
where the PhC is two-dimensional with a triangular array of holes
in an InP/GaInAsP dielectric background with a refractive index of
$n=3.24$. The radius of the air holes is $r=0.37\Lambda$. In order
to realize a non-solid waveguide, the PhC waveguide is formed by
slightly enlarging the radius ($r_1$) of the air holes in a single
row which are subsequently utilized for liquid-infiltration. In
order to obtain a high filling factor, we optimize the waveguide
structure and obtain $f\approx 40\%$ when $r_1=0.5\Lambda$. The
enhancement factor exceeds unity near the band edge, which is
attributed to both slow-light propagation as well as to the large
filling factor.

\section{Conclusion}

In conclusion, we have studied the potential of using
liquid-infiltrated photonic crystals to enhance optical
absorbance. Using a scattering-matrix approach we have re-derived
a general expression for the enhancement factor, thus confirming
the interaction-time interpretation offered in previous work. Our
results clearly demonstrate how optical absorbance may benefit
from slow-light phenomena combined with a filling factor close to
unity. With the aid of photonic crystals, it is possible to design
an optical enhanced detection system which has both a slow group
velocity and a high filling factor. The slow-light enhancement of
the absorption, by possibly an order of magnitude, may be traded
for yet smaller miniaturized systems or for increased sensitivity
of existing devices.

\begin{acknowledgement}
We thank K.~B. Mogensen and J.~P. Kutter for stimulating
discussion. This work is financially supported by the \emph{Danish
Council for Strategic Research} through the \emph{Strategic
Program for Young Researchers} (grant no: 2117-05-0037).
\end{acknowledgement}

%\bibliographystyle{OQE}
%\bibliographystyle{springer}
%\bibliography{Q:/papers/BibTeX/OFTS}

\end{document}